# Adverse weather amplifies social media activity

Kelton Minor*    Esteban Moro†    Nick Obradovich‡

**Abstract**

Humanity spends an increasing proportion of its time interacting online. Scholars are intensively investigating the societal drivers and resultant impacts of this collective shift in our allocation of time and attention. Yet, the external factors that regularly shape online behavior remain markedly understudied. Do environmental factors alter rates of online activity? Here we show that adverse meteorological conditions markedly increase social media use in the United States. To do so, we employ climate econometric methods alongside over three and a half billion social media posts from tens of millions of individuals from both Facebook and Twitter between 2009 and 2016. We find that more extreme temperatures and added precipitation each independently amplify social media activity. Weather that is adverse on both the temperature and precipitation dimensions produces markedly larger increases in social media activity. On average across both platforms, compared to the temperate weather baseline, days colder than -5°C with 1.5-2cm of precipitation elevate social media activity by 35%. This effect is nearly three times the typical increase in social media activity observed on New Year's Eve in New York City. We observe meteorological effects on social media participation at both the aggregate and individual level, even accounting for individual-specific, temporal, and location-specific potential confounds.

## Introduction

Social media is used by more than half of humanity, over nine in ten internet users, and seven in ten Americans[1,2]. Although social media platforms have been espoused for their capacity to boost social capital[3–5], they have also been engineered to capture human attention and engagement[6–10]. Over ten percent of the United States population spent over four hours per day on social media in 2019[11].

The ubiquity of social media has profoundly altered the way humans communicate, socialize, and coordinate. For example, social media can facilitate conversations about important issues in public health and public policy by incorporating voices from large segments of the global population[12]. Further, the immediacy of social media facilitates quick access to products and services[13], disaster awareness and response[14], crowdfunding[15], and rapid access to news and information[16].

However, social media has exacerbated the risk of cyber-bullying and harassment[17], has likely sped the spread of questionable information[18], and has reduced privacy and data security[19]. Consistent with the social displacement hypothesis[20], higher social media use is associated with decreased in-person social interaction with close contacts[7,21], and US adolescents in 2016 spent one hour less per day engaged in in-person social interactions compared to the pre-social media 1980s cohort[22].

---

*Data Science Institute, Columbia University

†Department of Mathematics and GISC, Universidad Carlos III de Madrid, Madrid, Spain and Media Lab, Massachusetts Institute of Technology

‡Project Regeneration and MIT Center for Real Estate, Sustainable Urbanization Lab. Corresponding authors, kelton.minor@columbia.edu and nobradov@mit.edu

Importantly, evidence suggests the welfare effects of offline and online socialization diverge in sign: reported mental health is positively associated with offline interaction but negatively associated with logged social media use[23]. While evidence suggests that social media can increase news consumption and group coordination,[5,24] studies also show that social media use can causally degrade mental health[21,25–30], can be habitually addictive[31], can worsen performance attention deficits[25], and can promote online activity while reducing engagement in healthier tasks[21,30].

Scholars have uncovered many manners in which external environmental factors can shape the nature of online behaviors among those already online. For example, social media post content can provide high resolution revealed cues for natural hazard detection and damage assessment[32–38], recording pollution impacts[39–41], and registering social responses to heatwaves and anthropogenic environmental disasters[42–45]. Further, prior research demonstrates that diurnal, seasonal and meteorological fluctuations can modify the nature of human lexical expressions on social media[52].

Much is known about how environmental conditions shape social media activities once individuals are already online. Yet – given the importance of social media activities to human welfare – surprisingly little is known about how external conditions influence participation in social media. And it remains a fundamental question whether such social media participation – in and of itself – is sensitive to environmental conditions.

Here we investigate the causal effects of meteorological conditions on participation in social media activities. To do this, we employ over three and a half billion social media posts from tens of millions of Americans across both Facebook and Twitter between 2009 and 2016 coupled with high resolution local meteorological data spanning the contiguous United States.

## Results

Using these data (see Figure 1 and *Methods*) and methods drawn from climate econometrics[53], we examine five primary questions:

First, does the weather outside alter the volume of social media activity online? Second, do the effects of temperature and precipitation alter social media activity in independent or compound manners? Third, are changes in social media activity driven predominantly by alterations to weather-related posting or are they observed in non-weather related posting, too? Fourth, do the aggregate city-level effects persist when examined within person over time? Fifth, how does the magnitude of the effects of the weather on social media activity compare to the size of the effects produced by other salient societal events?

**Marginal effects of temperature and precipitation**

To investigate our first question – if weather alters the volume of social media activity online – we combine our aggregated city-level post counts with our daily meteorological data (see *Methods*). We empirically model this relationship as:

$$ln(Y_{jmt}) = f(tmax_{jmt}) + g(precip_{jmt}) + h(\mu) + \gamma_t + \nu_{jm} + \epsilon_{jmt} \quad (1)$$

In the longitudinal (panel) model represented in Equation 1, *j* indexes cities, *m* indexes unique year-months, and *t* indexes day-of-study. Our dependent variable $ln(Y_{jmt})$ represents the natural log of our city-level daily post count.

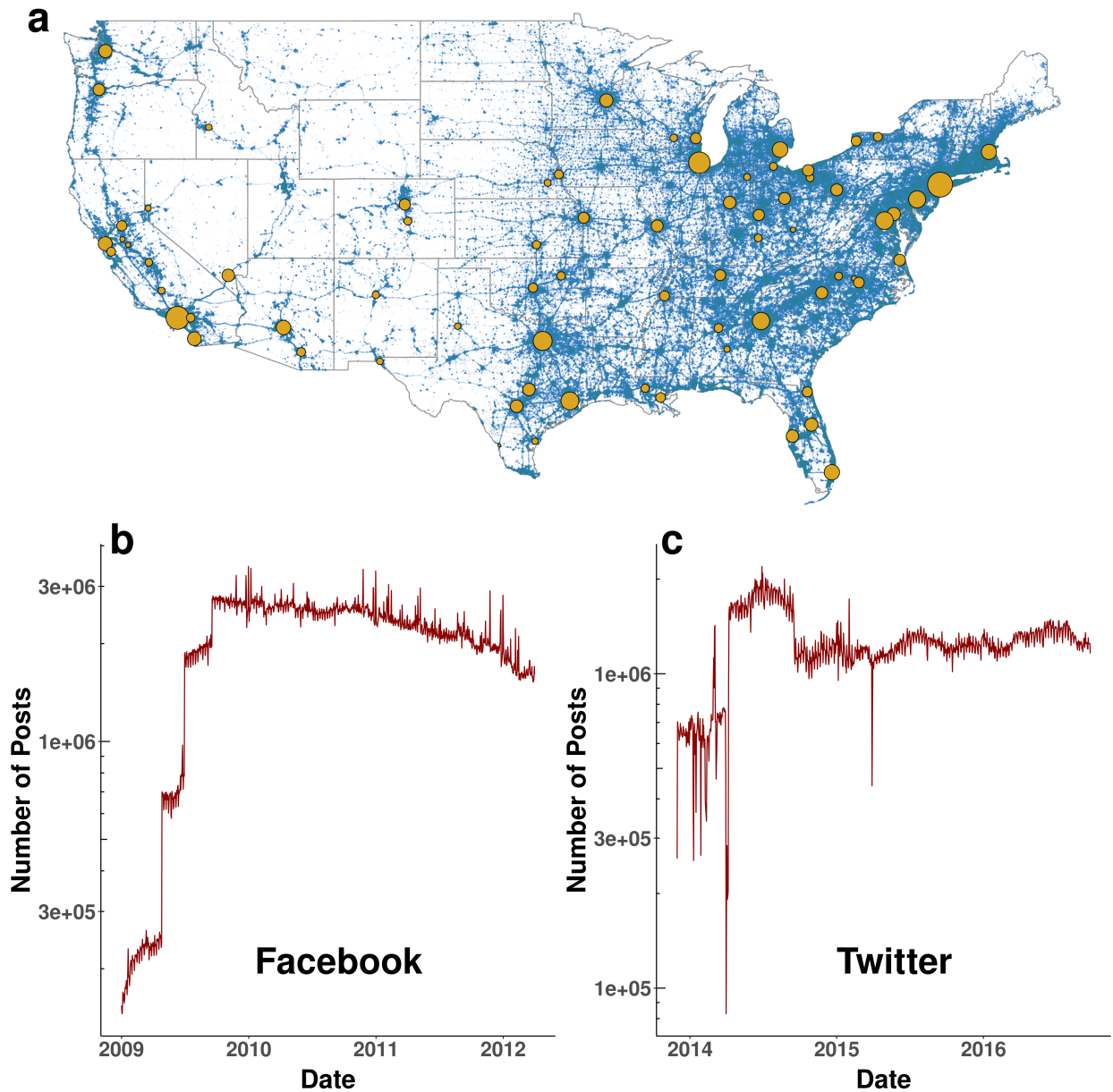

Figure 1. **Geographic location and temporal duration of social media data.** This figure depicts the US locales covered by our sources of social media data as well as the national daily variation in each series. Panel a plots the cross-sectional city-level variation of the social media data, with blue points indicating the location of geolocated tweet data and yellow points denoting the locale of cities in our analysis. Panel b displays the over-time variation in our Facebook data. The decrease in number of posts over time is due to changes in the Facebook platform over those years. Panel c depicts temporal variation in our Twitter data. The decline in number of posts in late 2014 is due to changes Twitter implemented in their geolocation process at that time.

Our focal independent variables in this analysis consist of daily maximum temperatures ($tmax_{jmt}$) and total precipitation ($precip_{jmt}$). We also control for daily temperature range, percentage cloud cover, and relative humidity, represented here via $h(\mu)$. We estimate our relationships of interest using indicator variables for each 5°C maximum temperature and temperature range bin, for each 1cm precipitation bin, and for each 20 percentage point bin of cloud cover and relative humidity (represented here by $f()$, $g()$, and $h()$ respectively). This approach enables flexible estimation of the relationship between our meteorological variables and social media post counts[48,53–56].

Unobserved geographic and temporal factors may alter social media activity in a manner that correlates with meteorological conditions. For example, people may exhibit more or less social media activity on average in cities that have better mass transit infrastructure or on dates when they are likely to have more leisure time. Further, there may exist unobserved, city-specific trends, such as changes in amount of daylight throughout the year or evolution in city-level economic conditions over time, that influence the social media activity within a city. To ensure that these factors do not confound our estimates of the effect of weather variables on social media activity, we include in Equation 1 $\nu_{jm}$ and $\gamma_t$ to represent city-by-month-of-study and day-of-study indicator variables (fixed effects), respectively. These variables account for all potentially confounding, constant, unobserved characteristics for each city across its seasons and for each unique date in the data across cities[53,57,58]. The remaining variation in our weather variables is thus as good as randomly assigned to the remaining variation in social media activity[53]. To bias our estimation, a confounding series would need to systematically co-vary with both meteorological anomalies and social media activity anomalies but not itself be caused by those weather anomalies[53,58].

We adjust for within-city and within-day correlation in $\epsilon_{jmt}$ by employing heteroskedasticity-robust standard errors clustered on both city and day-of-study[59]. We omit non-climatic control variables from Equation 1 because of their potential to generate bias in our parameters of interest (such variables are termed "bad controls" in the climate econometrics literature as they can introduce a form of post-treatment bias)[53,60,61].

We omit the 15-20°C maximum temperature, the 0-5°C diurnal temperature range, 0cm precipitation, 0-20% cloud cover, and 40-60% humidity indicator variables when estimating Equation 1. Our exponentiated coefficient estimates[62] can be interpreted as the percentage change in social media posts produced by a particular weather observation range compared to these baseline categories.

We present the results of the estimation of Equation 1 in Figure 2. As can be seen in Figure 2 panels (a) and (d), compared to moderate temperatures (15-20°C), the effects of both freezing temperatures and hot temperatures increase social media use. Freezing temperatures produce a 4.46% increase in social media activity on Facebook (p: < 0.001) and a 5.84% increase on Twitter (p: < 0.001). Temperatures above 40°C increase activity by 3.34% on Facebook (p: < 0.001) and by 3.58% on Twitter (p: < 0.001). Further, as can be seen in Figure 2 panels (b) and (e), added precipitation increases social media activity across both samples. Compared to the no precipitation baseline, 3-4cm of daily precipitation produces a 2.93% increase in social media activity on Facebook (p: < 0.001) and a 4.44% increase on Twitter (p: < 0.001).

## Nonlinear effects in the interaction between temperature and precipitation

Thus both temperature and precipitation – when considered independently – alter social media activity significantly and substantively. But are these effects simply additive in e.g. cold and wet conditions, or do they compound to produce nonlinear effects on social media activity?

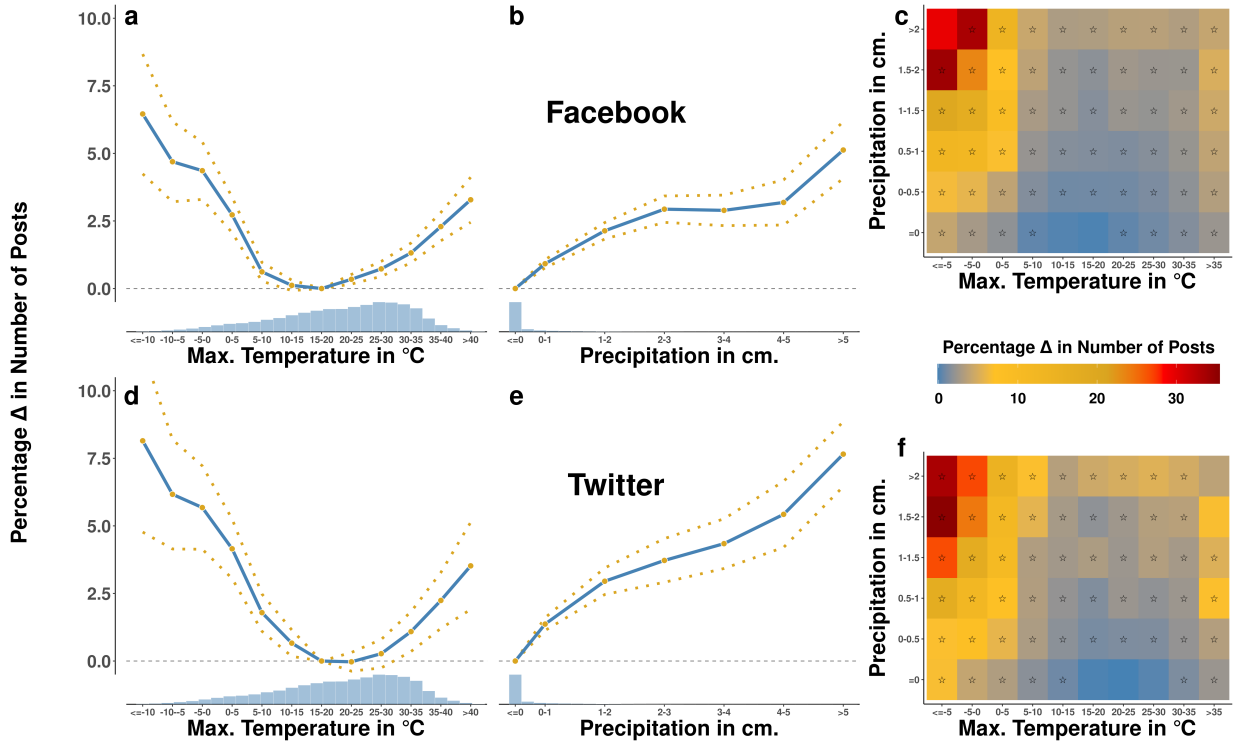

Figure 2. **Cold and hot temperatures, precipitation, and cold, wet conditions amplify number of social media posts.** Panel a displays the marginal effects estimated from our fixed effects regression model from Equation 1 of daily maximum temperatures on the log of the number of Facebook posts. Both colder and warmer temperatures amplify posting to Facebook relative to the 15-20°C reference range. Panel b depicts the marginal effect of precipitation on posting to Facebook. Greater amounts of daily precipitation amplify posting to the platform. Panel c depicts the effects associated with the interaction surface between temperature and precipitation on Facebook. Large and substantive increases in social media activity occur due to cold, wet temperatures. Panel d plots the marginal effects of daily temperature on posting to the Twitter platform. Panel e displays the effects of added daily precipitation on posting to Twitter. Panel f depicts the interaction surface between temperature and precipitation for the Twitter data. The functional forms of the marginal and interaction effects of temperature and precipitation are highly similar for both Facebook and Twitter, with effect sizes slightly larger for the effects of the weather on posting to Twitter. Shaded error bounds represent 95% confidence intervals calculated using heteroskedasticity-robust standard errors multiway clustered on both city and day-of-study. Stars in the cells in the interaction plots indicate that the 0.5th-99.5th percentile range of 1,000 cluster bootstrapped model estimates do not contain zero (see Methods).

To investigate this second question, we introduce what is to our knowledge a novel semi-parametric approach to estimating meteorological interaction surfaces in the context of climate econometrics. To estimate the semi-parametric interaction surfaces depicted in in Figure 2 (c) and (f), we estimate the model represented in Equation 2 separately for both the Facebook and Twitter data. This model is identical to Equation 1, but rather than estimate temperature and precipitation as only marginal entries into the model, we also estimate coefficients on the interaction of each meteorological bin.

$$ln(Y_{jmt}) = f(tmax_{jmt}) + g(precip_{jmt}) + f(tmax_{jmt}) * g(precip_{jmt}) + h(\mu) + \gamma_t + \nu_{jm} + \epsilon_{jmt} \quad (2)$$

In order to be identified, this model requires sufficient sample support under each bin. Because of the limited support on the extremes (for example there are very few observations that have both temperatures >40°C and >2 centimeters of precipitation), we limit the interaction surface in these models to 5°C temperature bin ranges from ($-Inf$-5°C]-(35°C-$Inf$) and half centimeter precipitation ranges from [0,0]-(2-$Inf$).

Using the coefficient estimates from Equation 2, we construct the simple effects for each temperature-precipitation bin[63]. This process omits as the reference bin the =0cm precipitation, (15°C-20°C] temperature interaction cell. To properly estimate the uncertainty in our estimates[63], we calculate median estimates and confidence regions for each simple effect by conducting 1,000 bootstrapped estimations of Equation 2, clustered by the city-level, and storing these 1,000 estimates for each grid cell in Figure 2(c) and (f). We then report the median estimate from this process in each cell and construct the 0.5th to 99.5th percentile range for each estimate[64]. If this confidence range does not include zero, we label that grid cell with a star in the figure.

The results of this estimation process uncover strong nonlinearity in the compound effects of temperature and precipitation on social media activity and can be seen in Figure 2(c) and (f). Compared to the mild weather baseline, conditions with temperatures below -5°C with 1.5-2cm of precipitation increase social media activity by 34.22% on Facebook and by 35.47% on Twitter. Precipitation during hot weather produces smaller – though still positive – effects. Compared to the mild weather baseline, temperatures above 35°C with 1-1.5cm of precipitation increase social media activity by 4.37% on Facebook and by 5.18% on Twitter (all results noted above are significant at the p < 0.01 level via cluster bootstrap inference).

**Weather-related and non-weather-related activity**

Our prior analyses examine changes to the volume of all types of posts within our data, inclusive of terms that may refer directly to the weather. It is possible that much of the increased activity observed in more adverse weather conditions relates only to added discussion of weather on the platforms. How much of a role does weather-related discussion play in our observed effects?

To classify weather-related posts in our sample, we employ a large crowd-sourced dictionary of terms (see *Methods* and previous work[48] for further details). We do not have access to the raw Facebook posts, so weather-term-related analyses are restricted to our Twitter data. Approximately 4% of tweets in our sample contained one or more of our weather terms.

To examine changes in the share of weather posting that result from changes in the weather, we modify Equation 1, substituting as the dependent variable the share of all tweets that weather-related tweets comprise on a given city-day. Otherwise estimation remains the same as in Equation 1.

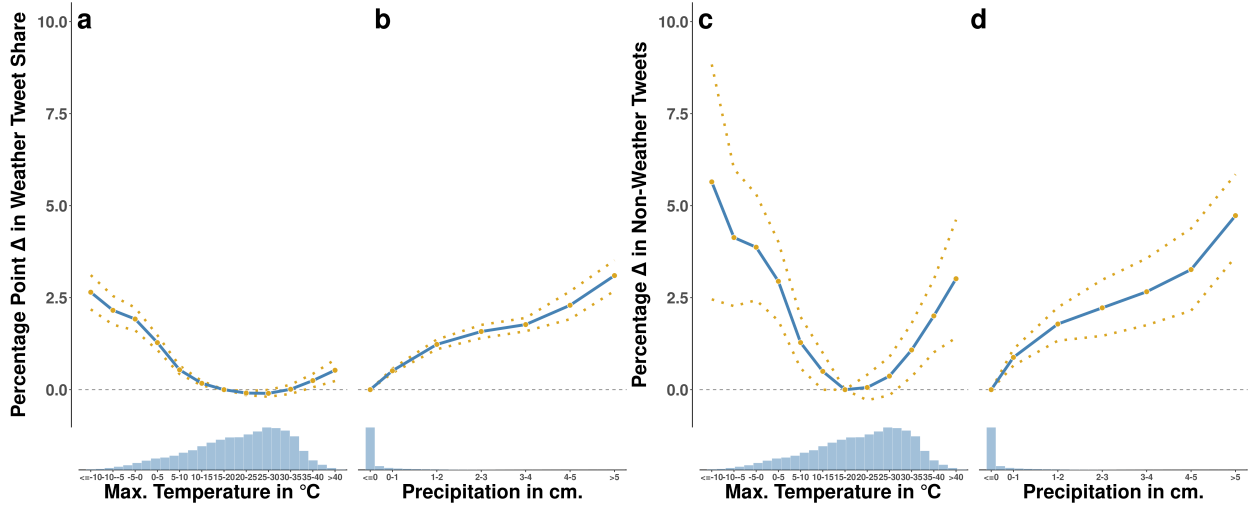

Figure 3. **Worse weather increases both weather-related and non-weather-related posting.** Overall changes in social media activity could be driven by individuals posting at much higher rates about the weather during more extreme weather conditions. Our Twitter data enable applying a crowdsourced definition of weather posts to examine changes in weather-related activity. Panels a and b indicate that the fraction of weather-related posts on Twitter notably increases with both cold temperatures and with added precipitation. However, panels c and d indicate that non-weather related posts also increase in more extreme conditions. Shaded error bounds represent 95% confidence intervals calculated using heteroskedasticity-robust standard errors multiway clustered on both city and day-of-study.

The results of this process can be seen in Figure 3, panels (a) and (b). More extreme temperatures and added precipitation both increase the share of weather-related tweets on Twitter. Freezing temperatures produce a 1.94 percentage point increase in the share of weather-tweets on the platform (p: < 0.001). Further, 3-4cm of daily precipitation produces a 1.78 percentage point increase in the share of weather-related Twitter posts (p: < 0.001).

While the share of weather-related posts on Twitter increases in more adverse conditions, so too does non-weather-related posting activity. To examine this, we again estimate Equation 1, but exclude from the sample any posts that are classified as weather-related by our classifier. The results – quite similar to those in the all-tweets analysis – can be seen in Figure 3, panels (c) and (d). Freezing temperatures produce a 3.95% increase in the share of non-weather posts on the platform (p: < 0.001). Further, 3-4cm of daily precipitation produces a 2.7% increase in the share of non-weather-related Twitter posts (p: < 0.001).

### Within-individual social media activity

Thus exposure to adverse weather significantly increases social media activity at the city level for both weather-related and non-weather related posting. However, effects observed at the city-level of aggregation may obscure sample composition dynamics that could produce problems of ecological inference. While cities may see more activity overall in worse weather conditions, this could be due to some individuals using the platforms in good weather and a different – larger – set of individuals using the platforms in adverse weather conditions. Do the same individuals, tracked over time, alter their social media participation in response to the weather?

For our individual-level analysis we employ user-day specific counts of posts on Twitter for a sample of individuals who authored messages on more than 25% of days in our sample, a sub-sample containing 2.17 million tweets across 366,855 individuals. For the purposes of computability, we take a simple random sample from this larger sampling frame of individuals to create a panel of 10,000 individuals representative of the frequent users in our Twitter data. We restrict our individual-level analysis to our Twitter data as we do not have access to the individual-level Facebook data.

To investigate if our city-level results persist within the same individuals over time, we employ these downsampled individual-level Twitter data, along with slight modifications to Equation 1 to estimate an individual fixed-effects empirical model. We estimate our individual-level relationship as:

$$ln(Y_{ijmt}) = f(tmax_{ijmt}) + g(precip_{ijmt}) + h(\mu) + \eta_i + \gamma_t + \nu_{jm} + \epsilon_{ijmt} \qquad (3)$$

In Equation 3, $i$ now indexes unique individuals and $\eta_i$ replaces $\alpha_j$ and represents individual-level indicator terms that control for individual-specific, time-invariant factors such as average propensity to participate in social media, constant individual demographic characteristics, as well as fixed weather preferences for each Twitter user in the sample[57]. The model again includes day-of-study and city-level by year-month indicator terms and estimation otherwise proceeds according to our city-level analysis.

As can be seen in Figure 4 panels (a) and (b), the effects of temperature and precipitation on within-individual posting activity mirror those that we observe in the city-level analysis. Compared to moderate temperatures (15-20°C), the effects of both freezing temperatures and hot temperatures increase individual social media use. Freezing temperatures produce a 3.19% increase in tweeting activity (p: 0.002), while temperatures above 40°C increase activity by 3.67% (p: < 0.001). And compared to the no precipitation baseline, 3-4cm of daily precipitation produces a 2.41% increase in tweet activity (p: 0.003).

**Effect sizes in context**

While the effects of the weather on social media activity thus persist within individual – even accounting for individual-level specific factors – how large are the effects we observe when compared to other significant factors that alter social media activity?

To understand the relative magnitude of the observed effects of adverse weather, we compare them to the increase in social media activity observed on a select set of large social events that occur in specific cities (see *Methods* for a comparison based on standard deviations in the deconvolved series). We again estimate Equation 2, with two modifications. First, we pool both the Facebook and Twitter data to generate an average treatment effect across these two sources. Second, in addition to the terms in Equation 2, we include indicator terms for each of the comparison events to estimate these parameters simultaneously alongside our meteorological variables. These events include the dates of the Boston Marathon in Boston, MA, the occurrences of New Year's Eve in New York City, NY, and the dates of Mardis Gras in New Orleans, LA. These indicator terms isolate the specific dates of each respective event, so that they are not collinear with the fixed effects in our models. For example, for the effect size of Mardis Gras on social media activity in New Orleans, our indicator variable is equal to one on each historical date of Mardis Gras for only those observations that fall within the city of New Orleans on those dates.[43,48]

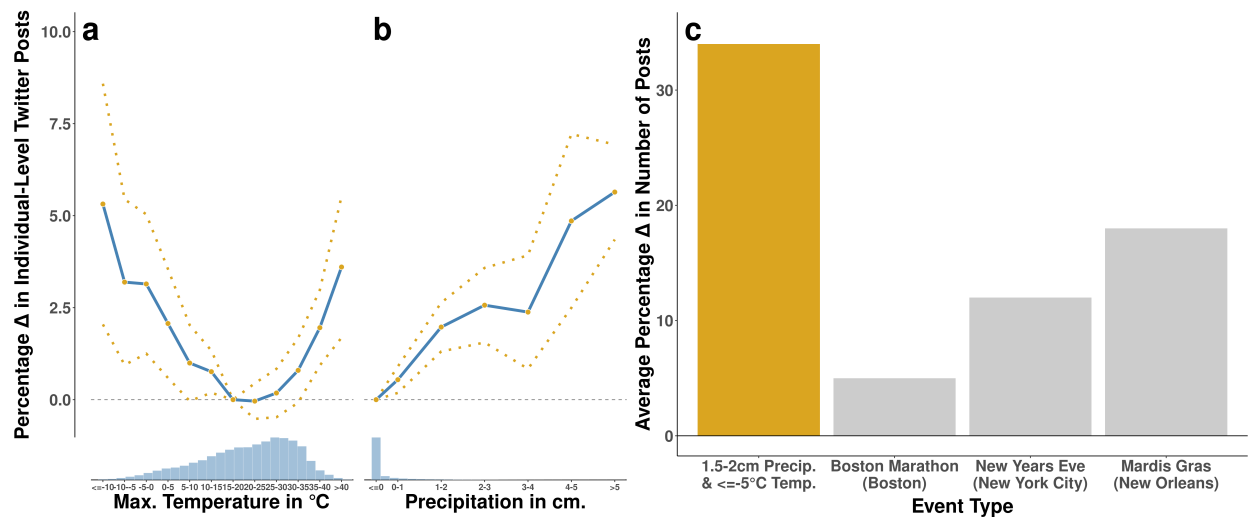

Figure 4. **Individual-level results and effect size comparisons.** Panels (a) and (b) conduct a individual-level regression – employing individual fixed effects – on a randomly sampled subset of 10,000 Twitter users active on more than 25% of days in our sample. As can be seen, similar effects are observed within individuals, indicating our results are not purely driven by changes in sample composition due to altered weather conditions. Shaded error bounds represent 95% confidence intervals calculated using heteroskedasticity-robust standard errors multiway clustered on both city and day-of-study. Panel (c) compares the pooled average effect size for both the Facebook and Twitter impacts of adverse weather conditions (less than -5°C and 1.5-2cm of precipitation) to the average effect of other events in our data. Adverse weather conditions increase social media activity by 34%, which is approximately three times the typical increase in activity on New Year's Eve in New York City. All effects in (c) are significantly different from zero at the $p < 0.01$ level.

Figure 4 panel (c) indicates that each of these events is significantly associated with increased social media use at the p < 0.01 level. A day of temperatures below -5°C and precipitation of between 1.5-2cm has an effect that is both statistically significant and substantively quite large. This effect of adverse weather (34%) is over six times the effect of the Boston Marathon in Boston (5%), nearly three times the effect size associated with New Year's Eve in New York City (12%), and nearly double the effect of Mardis Gras in New Orleans (18%). As it compares to other well-known and large-scale social events, adverse weather's impact on social media activity is quite large.

## Discussion

Over four billion people now use social media, yet the influence of environmental conditions on humanity's dominant mode of digital connection has remained unstudied[65,66]. Drawing on billions of posts from two popular social media platforms, including the largest in the world (Facebook), we find empirical support for a causal effect of adverse weather on social media use, with both hot and cold temperatures and precipitation increasing participation in social media. Further, we identify similar, non-linear social media responses to meteorological conditions for both Facebook and Twitter and show that compound weather events induce large magnitude increases in online social activity.

Compound extreme cold and heavy precipitation events boost local social media participation by considerably more than the Boston Marathon, Mardis Gras in New Orleans, and even New Year's Eve in New York City. Consistent meteorological effects on social media activity are evident at both the aggregate and individual level, adjusting for location-specific, seasonal, and time invariant between-person differences.

There are several important considerations relating to the findings of our study. First, by using data from both social media platforms, we take advantage of the relative strengths of each as a data source: Facebook data is more likely to be representative while Twitter data provides for comparison of results across social media contexts. By the end of our sampling period, nearly three quarters of online adults used Facebook, as compared to between 15 and 30% for Twitter, LinkedIn, Pinterest, and Instagram[67]. Surveyed adults also indicated that they use Facebook more frequently than any other social media platform, and similar proportions of the US population used Facebook and Twitter five years later[2]. Thus, even though our results suggest the effect of adverse weather on social media activity generalizes across platforms, we cannot rule out the possibility that other social media platforms may exhibit idiosyncratic responses to meteorological conditions that differ from those we find here.

Second, while we observe consistent weather-social media responses across years in our sample, social media platforms are in continuous flux, representing a moving target for researchers[68]. Thus, it remains unclear whether the results we identify here will generalize to far future versions of these social media platforms or alternative modes of online social engagement.

Third, although we find that the effects of meteorological factors on social media participation persist at the individual-level, approaches that directly monitor phone usage or that monitor cross-platform activity can enable more detailed decomposition of environmentally-driven social media use and digital behavior. These are important avenues for future research.

Fourth, it is possible that the geolocated Twitter data is unrepresentative of Twitter more broadly, since such data constitute a subset of all tweets. However, the Facebook data reflects a broader set of posting across the US. That the Facebook results mirror the Twitter results partially ameliorates this concern, suggesting that Twitter users are unlikely to be selectively registering opt-in geolocation as a function of weather.

Fifth, measurement error may exist between observed weather and the weather that users actually experience, possibly attenuating the magnitude of our estimates[69]. Thus the quite large effects we identify may actually be underestimates as compared to the effects that precise in-situ measurements of meteorology could produce. Issues of right-hand-side measurement error may also be particularly salient with respect to our measures of cloud cover and humidity, as they are derived from gridded reanalysis data rather than directly from sparse station observations[53].

Sixth, automated accounts in our data that circumvented bot detection filtering may bias our results if they were programmed to tweet as a direct function of adverse weather; more likely, such accounts might attenuate our effect estimates if their behaviors are invariant to meteorological conditions[70,71].

Finally, our analysis was conducted on those who self-select into social media use. Our results may not apply to demographics that are less likely to use either Facebook or Twitter, including older generations.

Ultimately, every human experiences the meteorological conditions where they live. As such, the weather's role in shaping the degree to which humans interact with one another in online settings – often mediated by social media platforms and algorithms – or offline settings is an important component of the scholarly attempt to characterize the external environmental and social factors that alter human social engagement[58,72–76]. While we uncover large effects of adverse weather on online social activity in this study, future studies are critically needed to provide broader insight into the suite of external factors that likely alter the degree to which humans interact with one another online versus offline.

# Methods

### Social media data

Our social media data are comprised of 3.5 billion total posts, with 2.4 billion Facebook posts and 1.1 billion Twitter posts. We employ data used in previous work (described in those works as well)[48,77]. These data are derived from underlying unique "status updates" which are natural language text-based messages that an individual's contacts may view on their own Facebook News Feed or Twitter Timeline.

Our Facebook data begin on January 1st, 2009 and end on March 31st, 2012, containing 1,176 days in total. The Facebook data consist of all individuals on the platform – both public and private accounts – that selected English as their language, chose the U.S. as their country of residence, and could be linked to our sample of metropolitan areas by their IP-based geographic location at time of their posting. The Facebook data we use here are also described in detail elsewhere[77].

Our Twitter data are comprised of individual posts, or "tweets", that are short messages limited to 140 characters (in the period under study) and are publicly viewable by others on the platform by default. Our Twitter data span from November 30th, 2013 to June 30th, 2016, including 938 days in the sample. We gathered tweets using Twitter's public Streaming API, placing a bounding box filter over the United States to gather the set of precisely geolocated tweets. We then assigned tweets falling within a metropolitan area's boundaries to that specific region. This procedure allows for a high level of certainty that each included tweet originated within a specific metropolitan area. We exclude retweets from our analysis and only consider direct user-generated content.

## Meteorological data

We use gridded (at ~4km) meteorological data from the PRISM Climate Group for our daily maximum temperature, temperature range, and precipitation variables[78]. We also employ cloud cover – a measure of sun exposure – and relative humidity data from the National Centers for Environmental Prediction (NCEP) Reanalysis II project[79]. We match daily meteorological variables in a location to the posts of individual social media users geolocated to that particular location on that day.

## Weather-related posts

We present our weather-term dictionary below:

>aerovane air airstream altocumulus altostratus anemometer anemometers anticyclone anticyclones arctic arid aridity atmosphere atmospheric autumn autumnal balmy baroclinic barometer barometers barometric blizzard blizzards blustering blustery blustery breeze breezes breezy brisk calm celsius chill chilled chillier chilliest chilly chinook cirrocumulus cirrostratus cirrus climate climates cloud cloudburst cloudbursts cloudier cloudiest clouds cloudy cold colder coldest condensation contrail contrails cool cooled cooling cools cumulonimbus cumulus cyclone cyclones damp damp damper damper dampest dampest degree degrees deluge dew dews dewy doppler downburst downbursts downdraft downdrafts downpour downpours dried drier dries driest drizzle drizzled drizzles drizzly drought droughts dry dryline fall farenheit flood flooded flooding floods flurries flurry fog fogbow fogbows fogged fogging foggy fogs forecast forecasted forecasting forecasts freeze freezes freezing frigid frost frostier frostiest frosts frosty froze frozen gale gales galoshes gust gusting gusts gusty haboob haboobs hail hailed hailing hails haze hazes hazy heat heated heating heats hoarfrost hot hotter hottest humid humidity hurricane hurricanes ice iced ices icing icy inclement landspout landspouts lightning lightnings macroburst macrobursts maelstrom mercury meteorologic meteorologist meteorologists meteorology microburst microbursts microclimate microclimates millibar millibars mist misted mists misty moist moisture monsoon monsoons mugginess muggy nexrad nippy NOAA nor'easter nor'easters noreaster noreasters overcast ozone parched parching pollen precipitate precipitated precipitates precipitating precipitation psychrometer radar rain rainboots rainbow rainbows raincoat raincoats rained rainfall rainier rainiest raining rains rainy sandstorm sandstorms scorcher scorching searing shower showering showers skiff sleet slicker slickers slush slushy smog smoggier smoggiest smoggy snow snowed snowier snowiest snowing snowmageddon snowpocalypse snows snowy spring sprinkle sprinkles sprinkling squall squalls squally storm stormed stormier stormiest storming storms stormy stratocumulus stratus subtropical summer summery sun sunnier sunniest sunny temperate temperature tempest thaw thawed thawing thaws thermometer thunder thundered thundering thunders thunderstorm thunderstorms tornadic tornado tornadoes tropical troposphere tsunami turbulent twister twisters typhoon typhoons umbrella umbrellas vane warm warmed warming warms warmth waterspout waterspouts weather wet wetter wettest wind windchill windchills windier windiest windspeed windy winter wintery wintry

**Standard-deviation comparison**

Another way to consider the magnitude of the meteorological effects is to compare them to the baseline variation in posting activity for the pooled Facebook and Twitter data once the fixed effects in Equation 1 or Equation 2 have been partialled out. To do so, we deconvolve our city-level data as in Equation 4 by regressing our logged number of daily city-level posts ($Y_{jmt}$) on the day-of-study ($\gamma_t$) and city-by-month-of-study ($\nu_{jm}$) fixed effects.

$$ln(Y_{jmt}) = \gamma_t + \nu_{jm} + \epsilon_{jmt} \quad (4)$$

This produces a residualized measure of the percentage change in daily posts that has had temporal factors and city-level trends removed. Comparing the variation of this deconvolved series to the magnitude of the effects estimated by regressing this series upon our deconvolved weather variables – as in Equation 2 – gives a sense of the relative size of the effects of the weather variables as compared to the baseline variation in the series once fixed effects have been partialled out.

Doing so, we find that the standard deviation of the percentage change in residualized daily posts is 7%. Thus the percentage change produced by temperatures below -5°C with precipitation of between 1.5-2cm (34%) represents a 5 standard deviation event.

## Acknowledgements

We thank Lorenzo Coviello, Haohui Chen, James H. Fowler, and Yury Kryvasheyeu for their assistance with the data underlying this work and thank Manuel Cebrian, Sune Lehman, and Iyad Rahwan for their comments.

## Author contributions

Conceptualization: N.O., E.M., K.M.; Data curation: N.O.; Formal analysis: N.O.; Investigation: N.O., K.M.; Methodology: N.O.; Software: N.O.; Visualization: N.O.; Writing, original draft: K.M., N.O.; Writing, review and editing: K.M., E.M., N.O.

## Competing interests

The authors declare no competing interests.

## Data availability

We collected our Twitter data from the public domain in adherence with Twitter's Developer Agreement and we used aggregated Facebook data published previously[48,77]. Intermediate data needed to evaluate the conclusions in the paper are present in the paper and are available from N.O. upon request. The raw social media data used in this study are restricted from public redistribution due to the terms of service of the respective platforms.

# References

1. Kemp, S. Digital 2021: Global Overview Report. *DataReportal – Global Digital Insights* (2021).

2. Auxier, B. & Anderson, M. Social Media Use in 2021. *Pew Research Center* (2021).

3. Ellison, N. B., Steinfield, C. & Lampe, C. The benefits of Facebook 'friends:' Social capital and college students' use of online social network sites. *Journal of computer-mediated communication* **12,** 1143–1168 (2007).

4. Appel, M., Marker, C. & Gnambs, T. Are social media ruining our lives? A review of meta-analytic evidence. *Review of General Psychology* **24,** 60–74 (2020).

5. Campante, F., Durante, R. & Tesei, A. Media and Social Capital. *Annual Review of Economics* **14,** 69–91 (2022).

6. Fogg, B. J. Persuasive technology: Using computers to change what we think and do. *Ubiquity* **2002,** 2 (2002).

7. Kuss, D. J. & Griffiths, M. D. Online social networking and addiction—a review of the psychological literature. *International journal of environmental research and public health* **8,** 3528–3552 (2011).

8. Kramer, A. D., Guillory, J. E. & Hancock, J. T. Experimental evidence of massive-scale emotional contagion through social networks. *Proceedings of the National Academy of Sciences* **111,** 8788–8790 (2014).

9. Zuboff, S. Big other: Surveillance capitalism and the prospects of an information civilization. *Journal of Information Technology* **30,** 75–89 (2015).

10. Crone, E. A. & Konijn, E. A. Media use and brain development during adolescence. *Nature communications* **9,** 1–10 (2018).

11. Clement, J. Number of social network users worldwide from 2017 to 2025. *Available online at https://www.statista.com/statistics/278414/number-of-worldwide-social-network-users/* **4,** 2020 (2020).

12. Castells, M. *Networks of outrage and hope: Social movements in the internet age.* (John Wiley & Sons, 2015).

13. Hajli, M. N. A study of the impact of social media on consumers. *International journal of market research* **56,** 387–404 (2014).

14. Kryvasheyeu, Y. *et al.* Rapid assessment of disaster damage using social media activity. *Science advances* **2,** e1500779 (2016).

15. Lu, C.-T., Xie, S., Kong, X. & Yu, P. S. Inferring the impacts of social media on crowdfunding. in *Proceedings of the 7th ACM international conference on web search and data mining* 573–582 (2014).

16. Matsa, K. E. & Walker, M. News consumption across social media in 2021. *Pew Research* (2021).

17. Hasebrink, U., Livingstone, S., Haddon, L. & Olafsson, K. *Comparing children's online opportunities and risks across europe: Cross-national comparisons for EU kids online.* (EU Kids Online, 2009).

18. Vosoughi, S., Roy, D. & Aral, S. The spread of true and false news online. *science* **359,** 1146–1151 (2018).

19. Garcia, D. Leaking privacy and shadow profiles in online social networks. *Science advances* **3,** e1701172 (2017).

20. Kraut, R. *et al.* Internet paradox: A social technology that reduces social involvement and psychological well-being? *American psychologist* **53,** 1017 (1998).

21. Allcott, H., Braghieri, L., Eichmeyer, S. & Gentzkow, M. The welfare effects of social media. *American Economic Review* **110,** 629–76 (2020).

22. Twenge, J. M., Spitzberg, B. H. & Campbell, W. K. Less in-person social interaction with peers among US adolescents in the 21st century and links to loneliness. *Journal of Social and Personal Relationships* **36,** 1892–1913 (2019).

23. Shakya, H. B. & Christakis, N. A. Association of Facebook use with compromised well-being: A longitudinal study. *American journal of epidemiology* **185,** 203–211 (2017).

24. Mosquera, R., Odunowo, M., McNamara, T., Guo, X. & Petrie, R. The economic effects of facebook. *Experimental Economics* **23,** 575–602 (2020).

25. Braghieri, L., Levy, R. & Makarin, A. Social Media and Mental Health. *American Economic Review* **112,** 3660–3693 (2022).

26. Hunt, M. G., Marx, R., Lipson, C. & Young, J. No more FOMO: Limiting social media decreases loneliness and depression. *Journal of Social and Clinical Psychology* **37,** 751–768 (2018).

27. Verduyn, P. *et al.* Passive Facebook usage undermines affective well-being: Experimental and longitudinal evidence. *Journal of Experimental Psychology: General* **144,** 480 (2015).

28. Sagioglou, C. & Greitemeyer, T. Facebook's emotional consequences: Why facebook causes a decrease in mood and why people still use it. *Computers in Human Behavior* **35,** 359–363 (2014).

29. Tromholt, M. The Facebook experiment: Quitting Facebook leads to higher levels of well-being. *Cyberpsychology, behavior, and social networking* **19,** 661–666 (2016).

30. Mosquera, R., Odunowo, M., McNamara, T., Guo, X. & Petrie, R. The economic effects of Facebook. *Experimental Economics* **23,** 575–602 (2020).

31. Allcott, H., Gentzkow, M. & Song, L. Digital Addiction. *American Economic Review* **112,** 2424–2463 (2022).

32. Sakaki, T., Okazaki, M. & Matsuo, Y. Earthquake shakes twitter users: Real-time event detection by social sensors. in *Proceedings of the 19th international conference on World wide web* 851–860 (2010).

33. Guan, X. & Chen, C. Using social media data to understand and assess disasters. *Natural hazards* **74,** 837–850 (2014).

34. Kryvasheyeu, Y. *et al.* Rapid assessment of disaster damage using social media activity. *Science advances* **2,** e1500779 (2016).

35. Spruce, M., Arthur, R. & Williams, H. Using social media to measure impacts of named storm events in the United Kingdom and Ireland. *Meteorological Applications* **27,** e1887 (2020).

36. Weaver, I. S., Williams, H. T. & Arthur, R. A social Beaufort scale to detect high winds using language in social media posts. *Scientific Reports* **11,** 1–13 (2021).

37. Arthur, R., Boulton, C. A., Shotton, H. & Williams, H. T. Social sensing of floods in the UK. *PloS one* **13,** e0189327 (2018).

38. Moore, F. C. & Obradovich, N. Using remarkability to define coastal flooding thresholds. *Nature communications* **11,** 1–8 (2020).

39. Jiang, W., Wang, Y., Tsou, M.-H. & Fu, X. Using social media to detect outdoor air pollution and monitor air quality index (AQI): A geo-targeted spatiotemporal analysis framework with Sina Weibo (Chinese Twitter). *PloS one* **10,** e0141185 (2015).

40. Zheng, S., Wang, J., Sun, C., Zhang, X. & Kahn, M. E. Air pollution lowers Chinese urbanites' expressed happiness on social media. *Nature Human Behaviour* **3,** 237–243 (2019).

41. Burke, M. *et al.* Exposures and behavioural responses to wildfire smoke. *Nature Human Behaviour* **6,** 1351–1361 (2022).

42. Romanello, M. *et al.* The 2021 report of the Lancet Countdown on health and climate change: Code red for a healthy future. *The Lancet* **398,** 1619–1662 (2021).

43. Romanello, M. *et al.* The 2022 report of the Lancet Countdown on health and climate change: Health at the mercy of fossil fuels. *The Lancet* **400,** 1619–1654 (2022).

44. Cody, E. M., Reagan, A. J., Mitchell, L., Dodds, P. S. & Danforth, C. M. Climate change sentiment on Twitter: An unsolicited public opinion poll. *PloS one* **10,** e0136092 (2015).

45. Ford, J. D. *et al.* Opinion: Big data has big potential for applications to climate change adaptation. *Proceedings of the National Academy of Sciences* **113,** 10729–10732 (2016).

46. Golder, S. A. & Macy, M. W. Diurnal and seasonal mood vary with work, sleep, and daylength across diverse cultures. *Science* **333,** 1878–1881 (2011).

47. Hannak, A. *et al.* Tweetin'in the rain: Exploring societal-scale effects of weather on mood. in *Proceedings of the International AAAI Conference on Web and Social Media* **6,** (2012).

48. Baylis, P. *et al.* Weather impacts expressed sentiment. *PloS one* **13,** e0195750 (2018).

49. Moore, F. C., Obradovich, N., Lehner, F. & Baylis, P. Rapidly declining remarkability of temperature anomalies may obscure public perception of climate change. *Proceedings of the National Academy of Sciences* **116,** 4905–4910 (2019).

50. Wang, J., Obradovich, N. & Zheng, S. A 43-million-person investigation into weather and expressed sentiment in a changing climate. *One Earth* **2,** 568–577 (2020).

51. Baylis, P. Temperature and temperament: Evidence from Twitter. *Journal of Public Economics* **184,** 104161 (2020).

52. Burke, M. *et al.* Higher temperatures increase suicide rates in the United States and Mexico. *Nature climate change* **8,** 723–729 (2018).

53. Hsiang, S. Climate Econometrics. *Annual Review of Resource Economics* **8,** 43–75 (2016).

54. Obradovich, N. Climate change may speed democratic turnover. *Climatic Change* **140,** 135–147 (2017).

55. Obradovich, N. & Fowler, J. H. Climate change may alter human physical activity patterns. *Nature Human Behaviour* **1,** 1–7 (2017).

56. Obradovich, N., Migliorini, R., Paulus, M. P. & Rahwan, I. Empirical evidence of mental health risks posed by climate change. *Proceedings of the National Academy of Sciences* **115,** 10953–10958 (2018).

57. Wooldridge, J. M. *Econometric analysis of cross section and panel data.* (MIT press, 2010).

58. Carleton, T. A. & Hsiang, S. M. Social and economic impacts of climate. *Science* **353,** aad9837 (2016).

59. Cameron, A. C., Gelbach, J. B. & Miller, D. L. Robust Inference With Multiway Clustering. *Journal of Business & Economic Statistics* **29,** 238–249 (2011).

60. Obradovich, N., Migliorini, R., Mednick, S. C. & Fowler, J. H. Nighttime temperature and human sleep loss in a changing climate. *Science Advances* **3,** e1601555 (2017).

61. Acharya, A., Blackwell, M. & Sen, M. Explaining Causal Findings Without Bias: Detecting and Assessing Direct Effects. *American Political Science Review* **110,** 512–529 (2016).

62. Giles, D. E. Interpreting Dummy Variables in Semi-logarithmic Regression Models: Exact Distributional Results. *University of Victoria Department of Economics Working Paper EWP* (2011).

63. Aiken, L. S., West, S. G. & Reno, R. R. *Multiple Regression: Testing and Interpreting Interactions.* (SAGE, 1991).

64. Good, P. I. *Permutation, Parametric, and Bootstrap Tests of Hypotheses.* (Springer Science & Business Media, 2006).

65. Creutzig, F. *et al.* Digitalization and the Anthropocene. *Annual Review of Environment and Resources* **47,** 479–509 (2022).

66. Stokols, D. *Social ecology in the digital age: Solving complex problems in a globalized world.* (Academic Press, 2018).

67. Greenwood, S., Perrin, A. & Duggan, M. Social Media Update 2016. *Pew Research Center* (2016).

68. Bayer, J. B., Trieu, P. & Ellison, N. B. Social media elements, ecologies, and effects. *Annual review of psychology* **71,** (2020).

69. Hausman, J. Mismeasured Variables in Econometric Analysis: Problems from the Right and Problems from the Left. *Journal of Economic Perspectives* **15,** 57–67 (2001).

70. Stella, M., Ferrara, E. & De Domenico, M. Bots increase exposure to negative and inflammatory content in online social systems. *Proceedings of the National Academy of Sciences* **115,** 12435–12440 (2018).

71. Shao, C. *et al.* The spread of low-credibility content by social bots. *Nature communications* **9,** 1–9 (2018).

72. Evans, G. W. Projected behavioral impacts of global climate change. *Annual review of psychology* **70,** 449–474 (2019).

73. Arcaya, M., Raker, E. J. & Waters, M. C. The social consequences of disasters: Individual and community change. *Annual Review of Sociology* **46,** 671–691 (2020).

74. Dietz, T., Shwom, R. L. & Whitley, C. T. Climate change and society. *Annual Review of Sociology* **46,** 135–158 (2020).

75. Klinenberg, E., Araos, M. & Koslov, L. Sociology and the Climate Crisis. *Annual Review of Sociology* **46,** 649–669 (2020).

76. Dube, O., Blumenstock, J. & Callen, M. Measuring Religion from Behavior: Climate Shocks and Religious Adherence in Afghanistan. (2022). doi:10.3386/w30694

77. Coviello, L. *et al.* Detecting emotional contagion in massive social networks. *PloS one* **9,** e90315 (2014).
78. Di Luzio, M., Johnson, G. L., Daly, C., Eischeid, J. K. & Arnold, J. G. Constructing Retrospective Gridded Daily Precipitation and Temperature Datasets for the Conterminous United States. *Journal of Applied Meteorology and Climatology* **47,** 475–497 (2008).
79. Kanamitsu, M. *et al.* NCEP–DOE AMIP-II Reanalysis (R-2). *Bulletin of the American Meteorological Society* **83,** 1631–1644 (2002).